\documentclass{ws-procs961x669}            
\newcommand{\ds}{\displaystyle}
\newcommand{\vev}[1]{\langle#1\rangle}
\newcommand{\mat}{\left ( \begin{array}}
\newcommand{\emat}{\end{array} \right )}
\newcommand{\vect}{\left ( \begin{array}{c}}
\newcommand{\evect}{\end{array} \right )}

\begin{document}
\title{Dulaity and charged pion condensation in chirally asymmetric dense quark matter 
 in the framework of an NJL$_2$ model}

\author{T. G. Khunjua, V. C. Zhukovsky}

\address{Faculty of Physics, Moscow State University,
119991, Moscow, Russia}

\author{
 K. G. Klimenko , R. N. Zhokhov }

\address{ State Research Center
of Russian Federation -- Institute for High Energy Physics,
NRC "Kurchatov Institute", 142281, Protvino, Moscow Region, Russia, zhokhovr@gmail.com}

\begin{abstract}
In this talk we present investigation of the phase structure of a
(1+1)-dimensional quark model with four-quark interaction
and in the presence of baryon ($\mu_B$), isospin ($\mu_I$) and chiral isospin ($\mu_{I5}$) chemical potentials. Spatially homogeneous and inhomogeneous (chiral density wave (for chiral condensate) and single wave (for charged pion condensate)) condensates are considered. It is established that in the large-$N_c$ limit ($N_c$ is the number of colored quarks) there exists a duality correspondence
between the chiral symmetry breaking phase and the charged pion
condensation (PC) one. 
The primary conclusion of this investigation is the fact that chiral isospin chemical potential generates charged pion condensation with non-zero baryon density in dense quark matter. 
Moreover, it is shown that inhomogeneous
charged PC phase with nonzero baryon density is induced in the model
by arbitrary small values of the chemical potential $\mu_{I5}$ (for a
rather large region of $\mu_B$ and $\mu_I$).
\end{abstract}


\bodymatter

\section{Introduction}

QCD at nonzero temperature and baryon chemical potential plays a fundamental role in
the description of a number of various physical systems. Two important ones are neutron stars, which probe the low temperature and intermediate baryon chemical
potential domain, and heavy ion collision experiments, which explore
the region of the high temperature and low baryon chemical potential
. However, the consideration of these systems is not possible in the framework of perturbative weak coupling QCD. Calculations with nonzero baryonic chemical potential $\mu_B$ is very hard to be performed on the lattice as well.

To describe physical situations, when the baryonic density is nonzero
 usually different effective theories are employed.
Among them, we especially would like to mention the NJL-type models \cite{njl}. 
They are nonrenormalizable in (3+1)-dimensional spacetime and can be considered only as effective field theories.
But there exist also low-dimensional theories, such as (1+1)-dimensional chiral Gross--Neveu (GN) type models \cite{gn,ft}, 
that possess a lot of common features with QCD (renormalizability, asymptotic freedom, dimensional transmutation, the spontaneous breaking of chiral symmetry) and can be used as a laboratory for the qualitative simulation of specific properties of QCD at {\it arbitrary energies}. 
 It is 
 well understood (see 
 \cite{barducci,chodos,thies}) that the usual {\it no-go} theorem \cite{coleman}, which generally forbids the spontaneous
breaking of any continuous symmetry in two-dimensional spacetime, does
not work in the limit  $N_c\to\infty$, where $N_c$ is the number of
colored quarks. 


Besides temperature and baryon density, there are additional
parameters 
, for instance, an isotopic chemical
potential $\mu_I$. It allows to consider systems with isospin
imbalance (different numbers of $u$ and $d$ quarks). It is realized, e.g.,
in neutron stars, heavy-ion experiments, etc. So QCD phase diagram in
the presence of both baryonic and isotopic chemical potentials has
been recently 
studied in 
 \cite{son,eklim,ak}, where the possibility of the
charged PC phase just at $\mu_I\ne 0$ was predicted.
However, the existence of the charged PC phase is established there
without sufficient certainty.
 Due to these circumstances, the question arises,
whether there exist any factors promoting the appearance of charged PC
phenomenon in dense baryonic matter.



 It was shown here in
the framework of toy NJL$_2$ model that this phase can be
realized if we take into account a nonzero chiral isotopic potential
in addition. This means that there should be chiral imbalance (a nonzero difference
between densities of left- and right-handed fermions) in the
system. Recall that chiral imbalance 
may arise from
the chiral anomaly in the quark-gluon-plasma phase of QCD and possibly
leads to the chiral magnetic effect \cite{fukus} in heavy-ion
collisions. It might be realized also in compact stars or condensed
matter systems \cite{andrianov} (-see also the review \cite{ms}).



The existence of spatially inhomogeneous phases in dense systems is certainly not a new idea. 
It is very challenging to
find inhomogeneous condensate as a solution and find its form analytically. However, more often one just assume some ansatz 
and then solve a minimax problem with respect to its parameters. 

In this paper we investigate the possibility of formation of homogneous and inhomogeneous
condensates in the system and
charged PC phenomenon in the framework of an extended (1+1)-dimensional NJL model with two quark flavors and in the presence of the baryon ($\mu_B$), isospin ($\mu_I$)  as well as chiral isospin ($\mu_{I5}$) chemical potentials. 
We will show that a chiral imbalance of dense and isotopically asymmetric baryon matter is a factor, which can induce there a charged
 PC phase.

Moreover, it has been shown in the framework of the NJL$_2$ model
under consideration that in the leading order of the large-$N_c$
approximation there arises a duality between chiral symmetry breaking
(CSB) and charged PC phenomena. 


\section{ The model and its thermodynamic potential}

We consider a two-dimensional model which is intended for simulation of the
properties of real dense quark matter with two massless quark flavors
($u$ and $d$ quarks). Its Lagrangian 
has the form
\begin{align}
& L=\bar q\Big [\gamma^\nu\mathrm{i}\partial_\nu
+\frac{\mu_B}{3}\gamma^0+\frac{\mu_I}2 \tau_3\gamma^0+\frac{\mu_{I5}}2
\tau_3\gamma^0\gamma^5\Big ]q+\frac{G}{N_c}\Big [(\bar qq)^2+(\bar q\mathrm{i}\gamma^5\vec\tau q)^2 \Big
],  \label{1}
\end{align}
where the quark field $q(x)\equiv q_{i\alpha}(x)$ is a flavor doublet ($i=1,2$ or $i=u,d$) and color $N_c$-plet ($\alpha=1,...,N_c$) as well as a two-component Dirac spinor (the summation in (\ref{1}) over flavor, color, and spinor indices is implied); $\tau_k$ ($k=1,2,3$) are Pauli matrices. The quantities $\gamma^\nu$ ($\nu =0,1$) and $\gamma^5$ in Eq. (1) are gamma matrices 
in (2+1)-dimensional space-time.
Baryon $\mu_B$, isospin $\mu_I$ and axial isospin $\mu_{I5}$ chemical potentials are introduced in order to describe quark matter with nonzero baryon $n_B$, isospin $n_I$ and axial isospin $n_{I5}$ densities, respectively. 

To find the thermodynamic potential of the system, we use a semi-bosonized version of the Lagrangian (\ref{1}) 
\begin{align}
&\widetilde L\ds =\bar q\Big [\gamma^\rho\mathrm{i}\partial_\rho
+\mu\gamma^0+ \nu\tau_3\gamma^0+\nu_{5}\tau_3\gamma^1-\sigma
-\mathrm{i}\gamma^5\pi_a\tau_a\Big ]q-\frac{N_c}{4G}\Big [\sigma\sigma+\pi_a\pi_a\Big ].
\label{2}
\end{align}
Starting from the theory (\ref{2}), one obtains in the leading order of the large $N_c$-expansion (i.e. in the one-fermion loop approximation) the following path integral expression for the effective action ${\cal S}_{\rm {eff}}(\sigma,\pi_a)$ of the bosonic $\sigma (x)$ and $\pi_a (x)$ fields:
$$
\exp(\mathrm{i}{\cal S}_{\rm {eff}}(\sigma,\pi_a))=
  N'\int[d\bar q][dq]\exp\Bigl(\mathrm{i}\int\widetilde L\,d^2 x\Bigr),
$$
 $N'$ is a normalization constant. 
The ground state expectation values  $\vev{\sigma(x)}$ and $\vev{\pi_a(x)}$ of the composite bosonic fields are determined by
the saddle point equations,
\begin{eqnarray}
\frac{\delta {\cal S}_{\rm {eff}}}{\delta\sigma (x)}=0,~~~~~
\frac{\delta {\cal S}_{\rm {eff}}}{\delta\pi_a (x)}=0,~~~~~
\label{05}
\end{eqnarray}

We use the following spatially inhomogeneous CDW ansatz for chiral condensate and the single plane wave ansatz for charged pion condensates:
\begin{align}
&\vev{\sigma(x)}=M\cos (2kx),~~~\vev{\pi_3(x)}=M\sin
(2kx),~~~\nonumber\\
&\vev{\pi_1(x)}=\Delta\cos(2k'x),~~~
\vev{\pi_2(x)}=\Delta\sin(2k'x), \label{06}
\end{align}
where gaps $M,\Delta$ and wavevectors $k,k'$ are constant dynamical quantities. 

In the leading order of the large $N_c$-expansion after performing the so-called Weinberg (or chiral) transformation it is possible to find for the TDP the following expression:
\begin{eqnarray}
\hspace{-0.3cm}\Omega (M,k,k',\Delta)~&& =\frac{M^2+\Delta^2}{4G}+\mathrm{i}\int\frac{d^2p}{(2\pi)^2}\ln
P_4(p_0).
\label{9}
\end{eqnarray}
In Eq. (\ref{9}) we use the notations
$P_4(p_0)=\epsilon_1\epsilon_2\epsilon_3\epsilon_4=\eta^4-2a\eta^2-b\eta+c,$
where $\eta=p_0+\mu$ and
\begin{eqnarray}
\hspace{-0.3cm}a&&=M^2+\Delta^2+p_1^2+\tilde\nu^2+\tilde\nu_{5}^2;~~b=8p_1\tilde\nu\tilde\nu_{5};\nonumber\\
\hspace{-0.3cm}c&&=a^2-4p_1^2(\tilde\nu^2+\tilde\nu_5^2)-4M^2\tilde\nu^2-4\Delta^2\tilde\nu_5^2-4\tilde\nu^2\tilde\nu_5^2.
\label{10}
\end{eqnarray}
It is evident that 
the TDP is an even function over each of the variables $M$ and $\Delta$ 
as well as over $\mu$,  $\tilde\nu$, $\tilde\nu_5$. 
Hence,
 we can consider in the following only $\mu\ge 0$, $\tilde\nu\ge 0$, $\tilde\nu_5\ge 0$, $M\ge 0$, and $\Delta\ge 0$ values of these quantities. Moreover, the expression (\ref{9}) for the TDP is invariant with respect to the so-called duality transformation,
\begin{eqnarray}
{\cal D}:~~~~M\longleftrightarrow \Delta,~~\nu\longleftrightarrow\nu_5,~~k\longleftrightarrow k'.
 \label{16}
\end{eqnarray}
It means that in the leading order of the large-$N_c$ approximation there is the so-called duality correspondence between chiral symmetry breaking (CSB) and charged PC phenomena. 



\subsection{ Thermodynamic potential }

It can be shown numerically that GMP of the TDP can never be of the form $(M_0\ne 0,\Delta_0\ne 0)$. Hence, in order to establish the phase portrait of the model, it is enough to study the projections $F_1(M)\equiv\Omega^{ren} (M,\Delta=0)$ and $F_2(\Delta)\equiv\Omega^{ren}(M=0,\Delta)$ of the TDP to the $M$ and $\Delta$ axes, correspondingly. 
It is possible to obtain the following expressions for these quantities in the case $k=0$, $k'=0$,
\begin{align}
&F_1(M)=\frac{M^2}{2\pi}\ln\left(\frac{M^2}{m^2}\right)
-\frac{M^2}{2\pi}-\frac{\nu_5^2}{\pi}-\theta (\mu+\nu-M)\frac{\cal A}{\pi}\nonumber\\
&-\theta
(|\mu-\nu|-M)\theta(\sqrt{(\mu-\nu)^2-M^2}-\nu_5)\frac{\cal B}{2\pi}\nonumber\\
&+\theta (\mu+\nu-M)\theta
  (\sqrt{(\mu+\nu)^2-M^2}-\nu_5)\frac{\cal C}{2\pi },\label{33}
\end{align}
where
\begin{align}
&{\cal A}=(\mu+\nu)\sqrt{(\mu+\nu)^2-M^2}
-M^2\ln\frac{\mu+\nu+\sqrt{(\mu+\nu)^2-M^2}}{M},
\end{align}
\begin{align}
&{\cal B}=|\mu-\nu|\sqrt{(\mu-\nu)^2-M^2}
+\nu_5\sqrt{\nu_5^2+M^2}\nonumber\\
&-2|\mu-\nu|\nu_5-M^2\ln\frac{|\mu-\nu|+\sqrt{|\mu-\nu|^2-M^2}}{\nu_5+\sqrt{\nu_5^2+M^2}},
\end{align}
\begin{align}
&{\cal C}=(\mu+\nu)\sqrt{(\mu+\nu)^2-M^2}
+\nu_5\sqrt{\nu_5^2+M^2}-2(\mu+\nu)\nu_5\nonumber\\
&-M^2\ln\frac{\mu+\nu+\sqrt{(\mu+\nu)^2-M^2}}{\nu_5+\sqrt{\nu_5^2+M^2}}.
\end{align}
\begin{equation}
F_2(\Delta)=F_1(\Delta)\Bigg |_{\nu\longleftrightarrow\nu_5}. \label{34}
\end{equation}



\begin{figure*}
\includegraphics[width=0.45\textwidth]{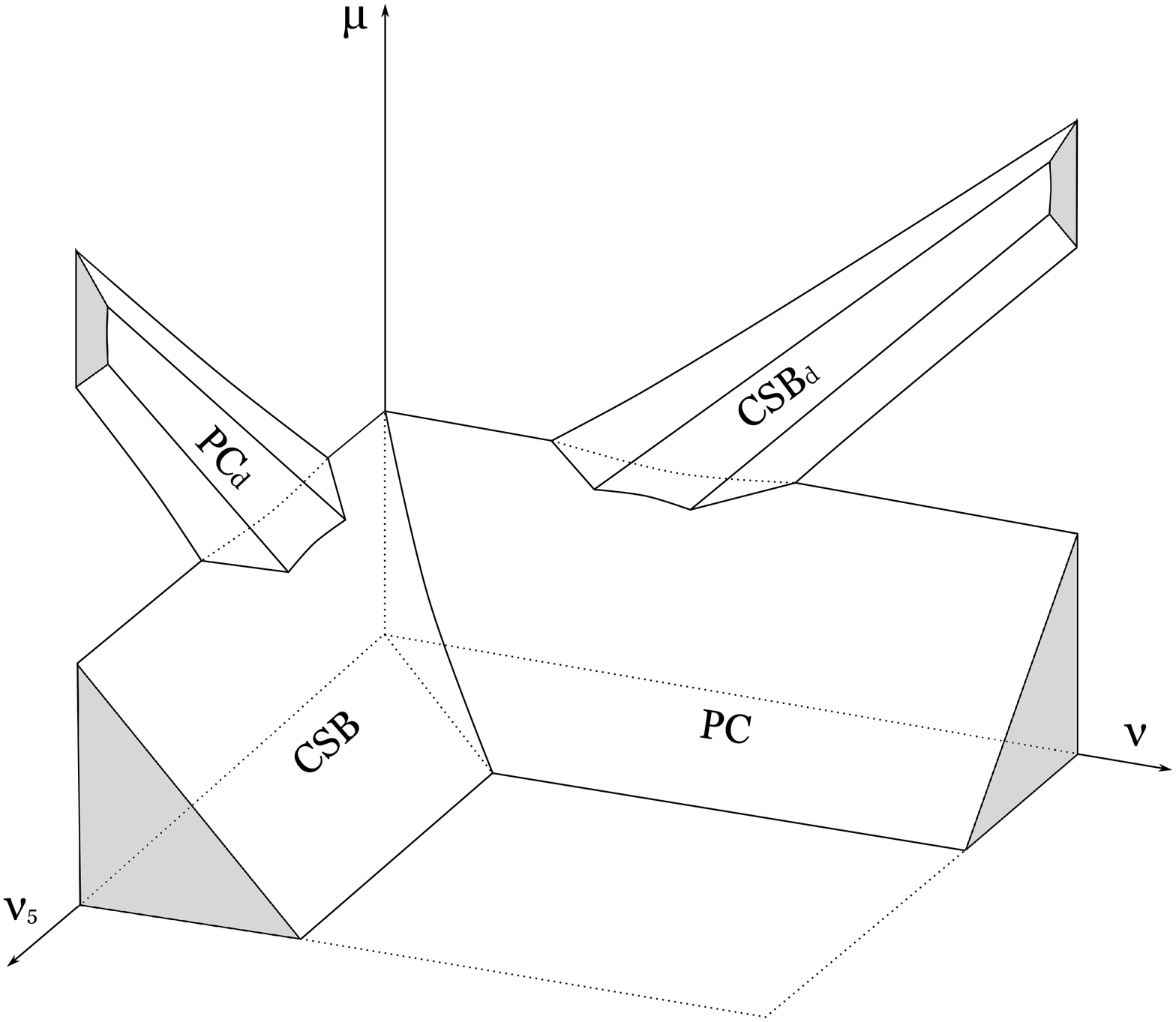}
\hfill
\includegraphics[width=0.45\textwidth]{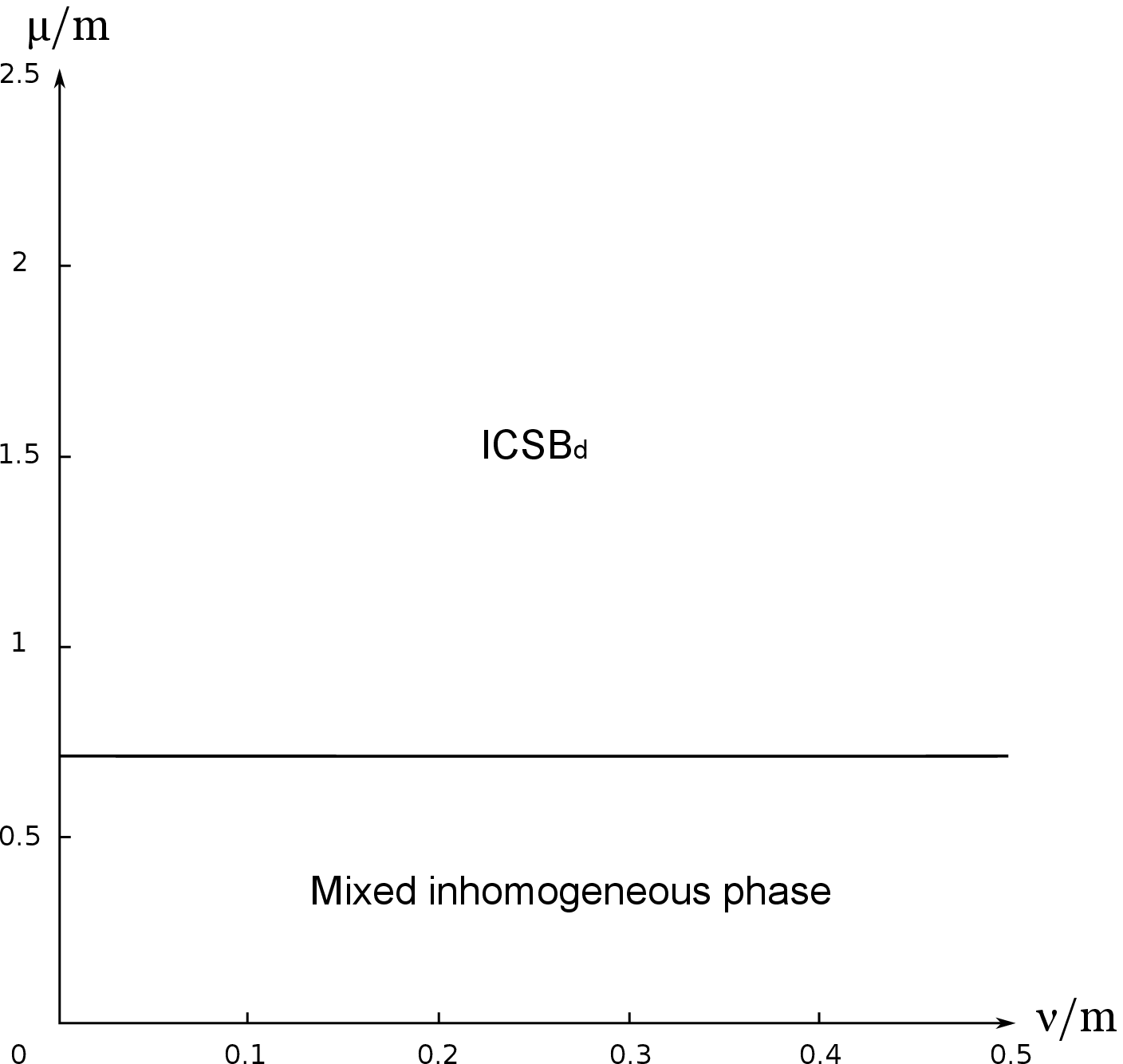}\\
\parbox[t]{0.45\textwidth}{\caption{Schematic representation of the $(\nu_5,\nu,\mu)$-phase portrait of the model in the case of spatially homogeneous condensates. 
}}\hfill
\parbox[t]{0.45\textwidth}{\caption{{\bf The case of 
inhomogeneous condensates}: The $(\nu,\mu)$-phase portrait at  $\nu_5=0_+$. 
}}
\end{figure*}



 The 
 TDP with $k,k'\ne 0$  $\Omega^{un}(M,k,k',\Delta)$ can be obtained from the 
 TDP in the 
 homogeneous case simply performing there the replacement $\nu, \nu_5\to\tilde{\nu}\equiv\nu+k, \tilde{\nu_5}\equiv\nu_5+k'$. 
However, the TDP  has several unphysical properties such as (i) the unboundedness from below with respect to the variables $k,k'$. (The unboundedness from below of the TDP is evident, e.g., from the expression (\ref{33}) if $\nu_5\to\tilde\nu_5$. 
(ii) Moreover, one can observe immediately that at $M=0$ and $\Delta=0$ the expression for the thermodynamic potential does depend on $k$ and  $k^{\prime}$. Bearing in mind the expression (\ref{06}) it is obvious that this is also quite unphysical and we need to change somehow the expression for thermodynamic potential in such a way that this dependence is eliminated.

The above mentioned nonphysical properties should
be eliminated by, e.g., the subtraction operation 
 applying
twice, first with respect to the variables $M,k$ and then with respect
to $\Delta,k'$. As a result, we have the
following physically relevant TDP,
\begin{align}
&\Omega^{phys}(M,k,k',\Delta)=\Omega^{ren}(M,k,k',\Delta)-\Omega^{ren}(M,k,k',0)+\Omega^{ren}(M,k,0,0) \nonumber\\
&-\Omega^{ren}(0,k,k',\Delta)+\Omega^{ren}(0,0,k',\Delta)-\Omega^{ren}(0,k,0,0)\nonumber\\
&-\Omega^{ren}(0,0,k',0)+\Omega^{ren}(0,k,k',0)+\Omega^{ren}(0,0,0,0).
\label{43}
\end{align}
It turns out that 
the mixed phase ($(M_0\ne 0,\Delta_0\ne 0)$) 
is absent in the model in inhomogeneous case as well, so it is enough to study only the projections of the TDP 
on the $M$ and $\Delta$ axes.


There could be several phases in the model (1). The first one is the symmetric phase 
 with zero gaps $M_0=0,\Delta_0=0$ and zero values of the wavevectors $k_{0}=0,k^{\prime}_{0}=0$. In the chiral symmetry breaking CSB (in the charged pion condensation CPC) phase 
 the TDP reaches the least value at the global minimum point with $M_0\ne 0,\Delta_0=0$ 
 or ($M_0=0,\Delta_0\ne 0$. 

 \subsection{ Phase diagrams and duality property of the model }
 In Fig. 1 the $(\mu,\nu,\nu_5)$-phase portrait of the model is presented in the supposition that all condensates are spatially homogeneous \cite{ekk}. It is clear from this figure that charged PC phase with nonzero quark number density $n_q$ (this phase is denoted there by PCd) can be realized in the model (1) only at rather large values of $\nu_5$. Now let us discuss the the case of spatially inhomogeneous condensates
\begin{figure*}
\includegraphics[width=0.45\textwidth]{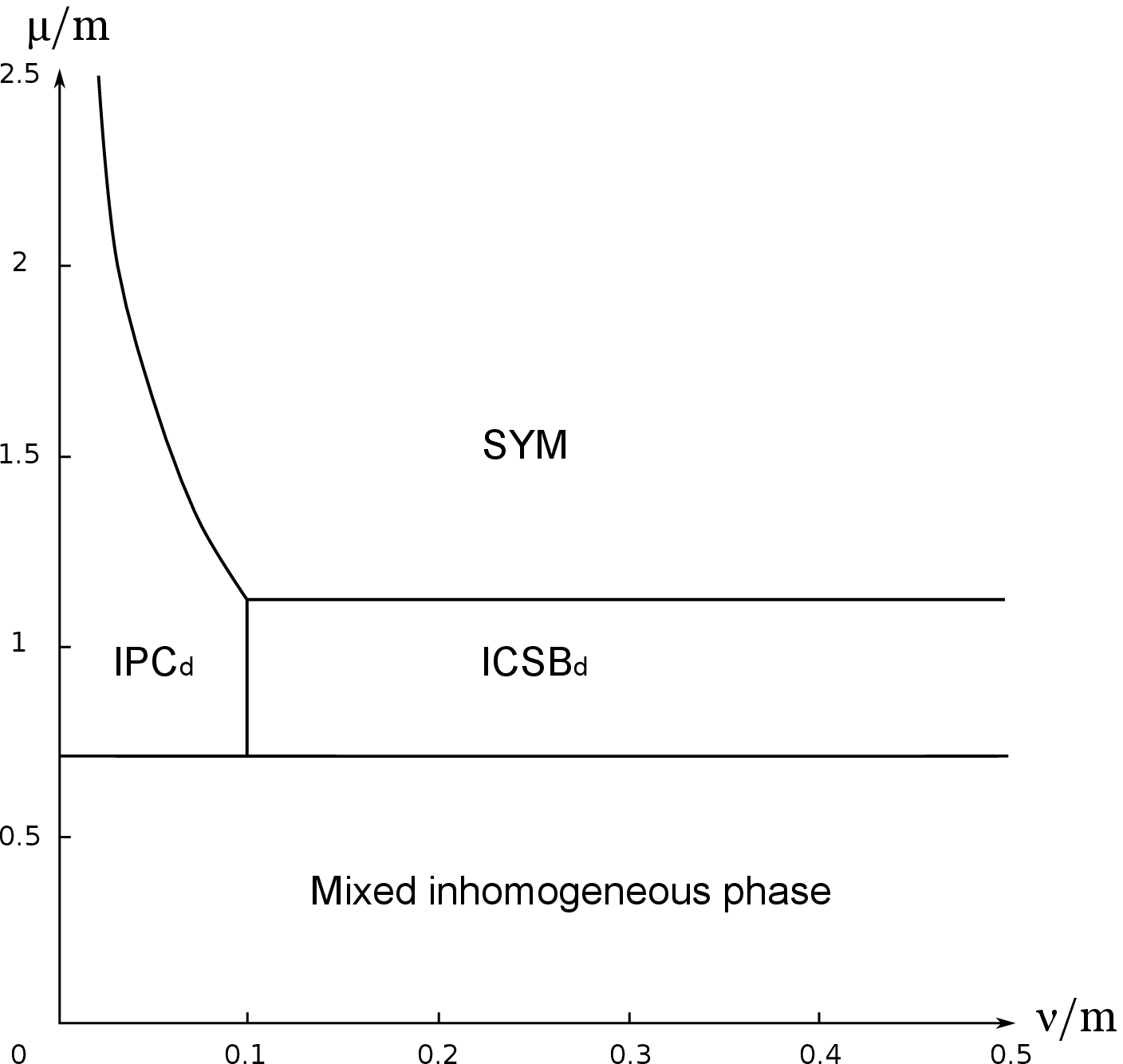}
\hfill
\includegraphics[width=0.45\textwidth]{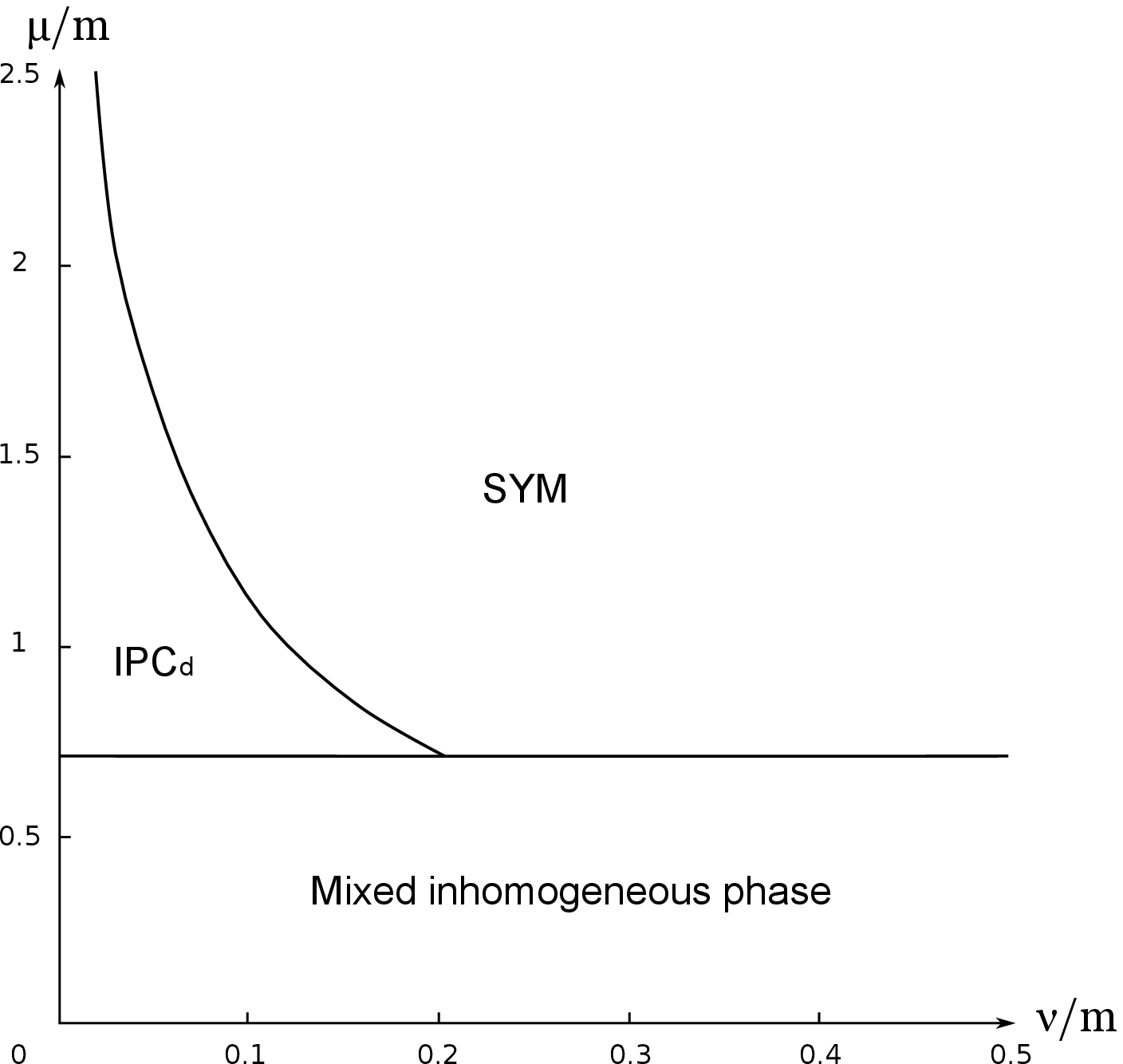}\\
\caption{{\bf The case of spatially inhomogeneous condensates}: The $(\nu,\mu)$-phase portrait of the model at  $\nu_5=0.1m$ (left figure) and at  $\nu_5\ge 0.2m$ (right figure). 
}
\end{figure*}



\subsubsection{Different $(\mu,\nu)$-phase diagrams}
\label{mixed}


 At infinitesimal values (see Fig. 2) of $\nu_5$ if one looks at the region of $\mu> m/\sqrt{2}$ one can see there 
 ICSBd phase (
symbol ''d`` means that quark number density is nonzero), at lower values of $\mu$ there is a region which is called ''Mixed inhomogeneous phase``. It turns out that for each point $(\mu,\nu)$ belonging to this region the TDP has two degenerate global minima, first of them 
corresponds to ICSB phase, the second 
to inhomogeneous charged pion condensation (IPC) phase. 
\begin{figure*}
\includegraphics[width=0.45\textwidth]{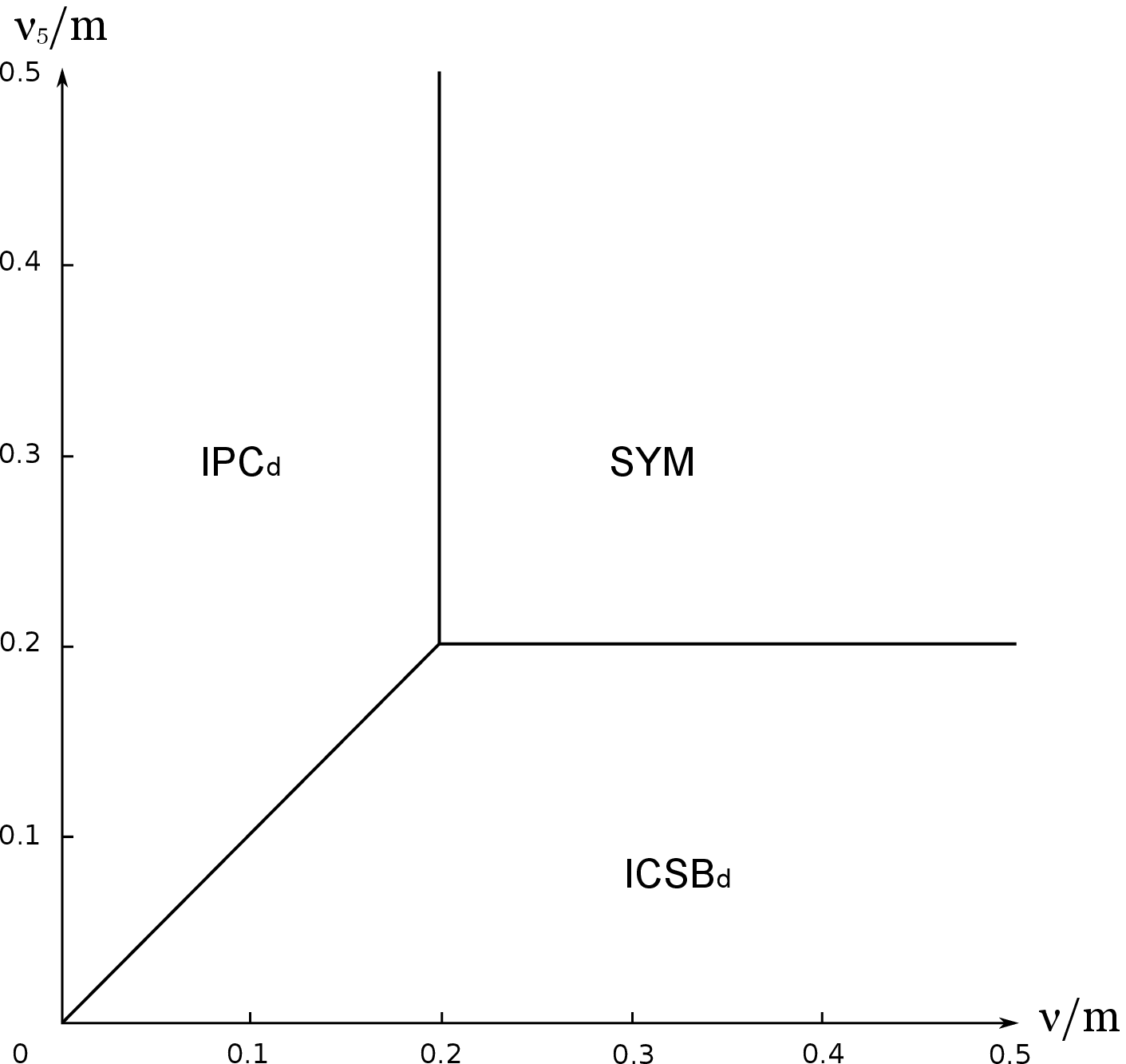}
\hfill
\includegraphics[width=0.45\textwidth]{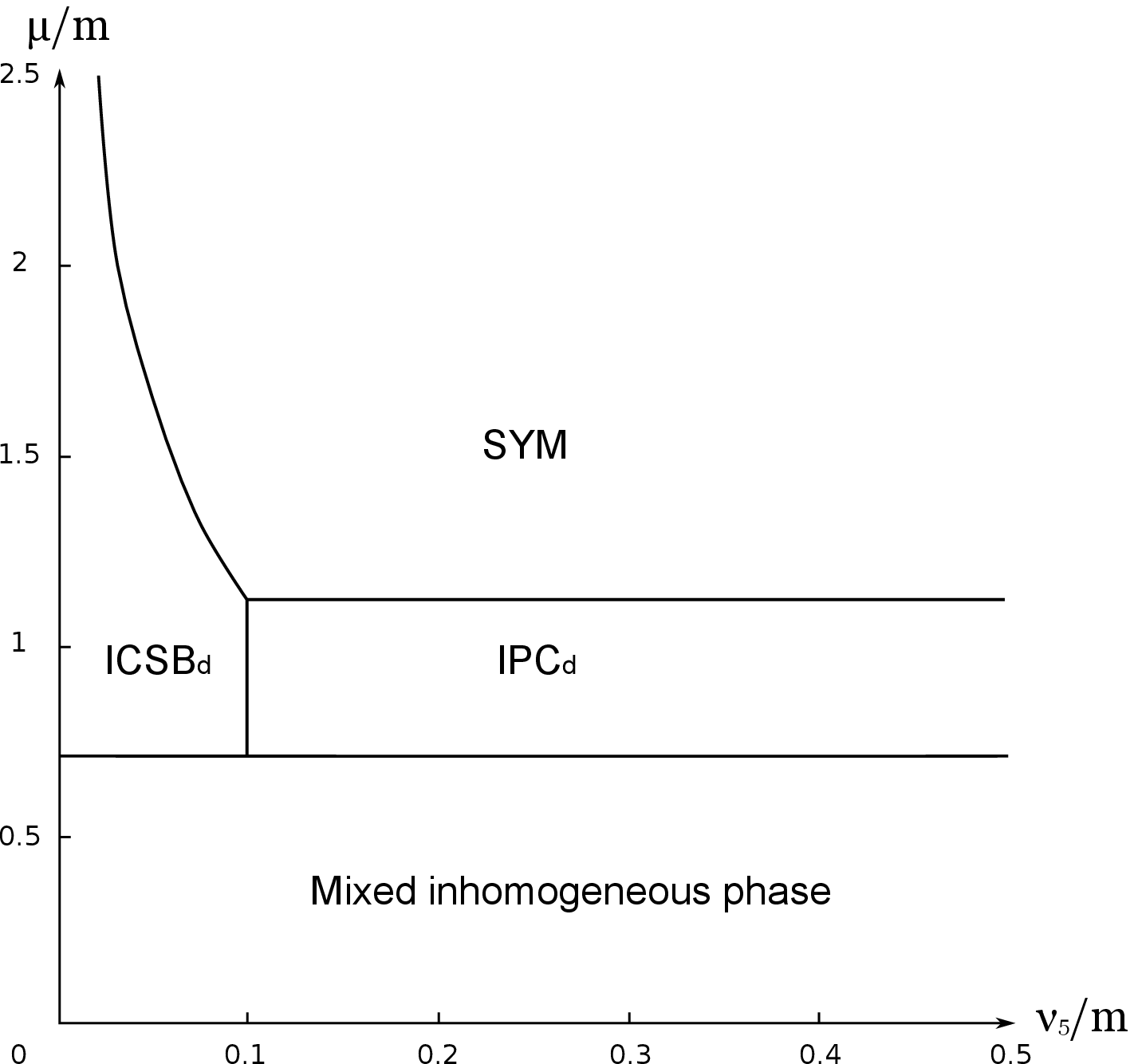}\\
\parbox[t]{0.45\textwidth}{\caption{{\bf The case of spatially inhomogeneous condensates}: The $(\nu,\nu_5)$-phase portrait of the model at  $\mu=0.75m$.
All notations are described in Figs 2, 3.}}\hfill
\parbox[t]{0.45\textwidth}{\caption{{\bf The case of spatially inhomogeneous condensates}: The $(\nu_5,\mu)$-phase portrait of the model at  $\nu=0.1m$. All notations are described in Figs 2, 3.}}
\end{figure*}

The structure of $(\mu,\nu)$-phase diagrams at other fixed values of the chiral chemical potential $\nu_5$ can be easily understood from the phase portraits of Figs 3, where $(\mu,\nu)$-phase diagrams are presented for two qualitatively different values of $\nu_5$. It is clear from the figure that at each finite $\nu_5>0$ the $(\mu,\nu)$-phase diagram contains 
IPCd phase. Moreover, the greater $\nu_5$, the smaller the size of the ICSBd phase, which disappears from a $(\mu,\nu)$-phase portrait at $\nu_5\ge 0.2m$. Hence, in
 the framework of the initial NJL$_2$ model, the chiral chemical potential $\nu_5$ serves as a factor, which promotes the charged pion condensation
 phenomenon in dense quark matter (it is the IPCd phase in all figures).
%

\subsubsection{Other phase diagrams and the role of duality }

 Let us discuss the role 
 of the duality 
 (\ref{16}) of the 
 TDP 
 on the phase structure.
Suppose that at some fixed 
$\mu$, $\nu=A$ and $\nu_5=B$ 
point 
CSB phase is
realized in the model. 
Then it follows from the duality 
of the TDP 
that at permuted chemical potential values (i.e. at $\nu=B$ and $\nu_5=A$ and 
the same $\mu$) 
PC phase is realized (and vice versa). This is the so-called duality correspondence between CSB and charged PC phases. 



Under the duality transformation 
the most general $(\nu,\nu_5,\mu)$-phase portrait is mapped to itself 
(self-dual). 
\begin{figure*}
\includegraphics[width=0.45\textwidth]{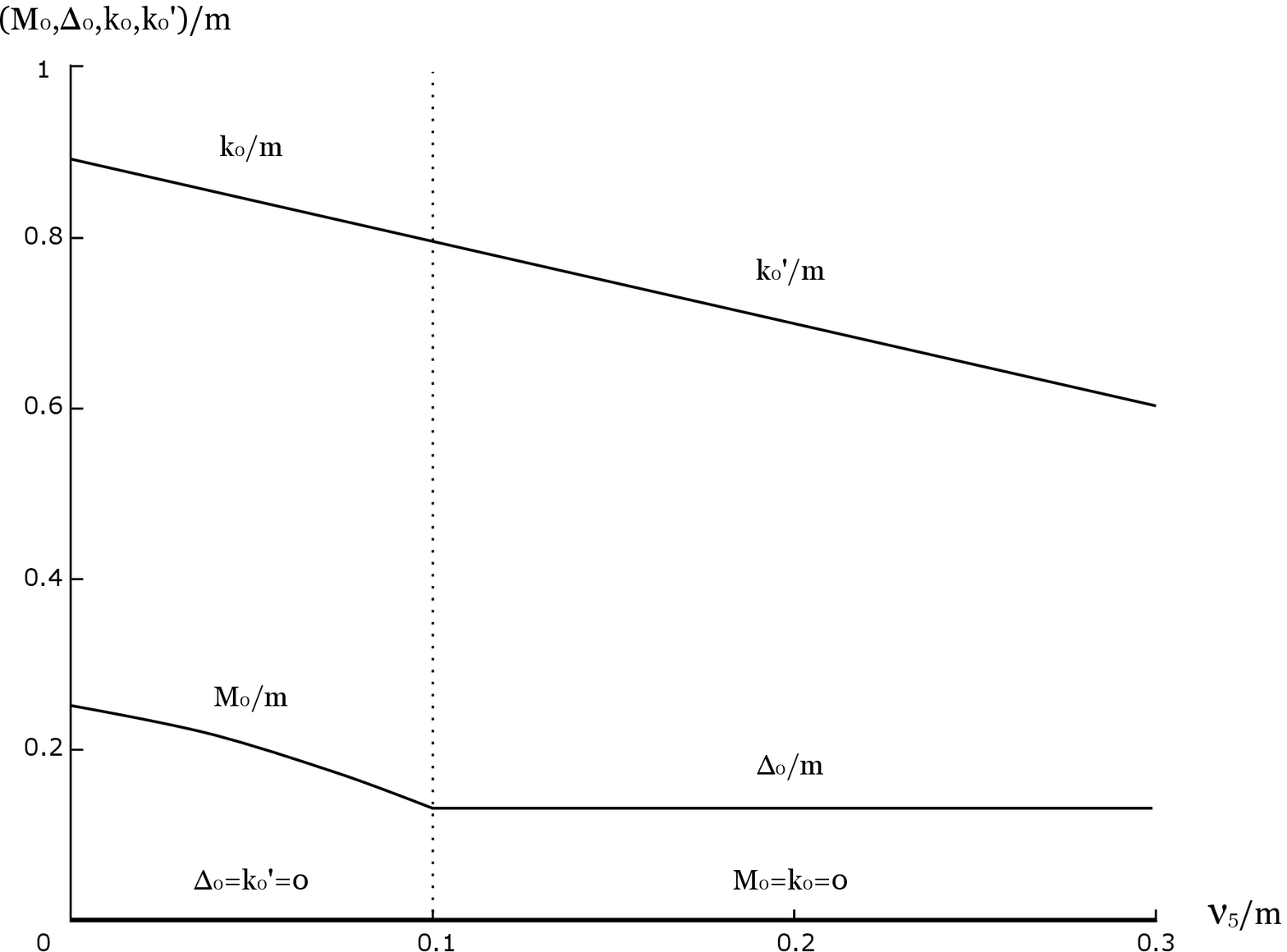}
\hfill
\includegraphics[width=0.45\textwidth]{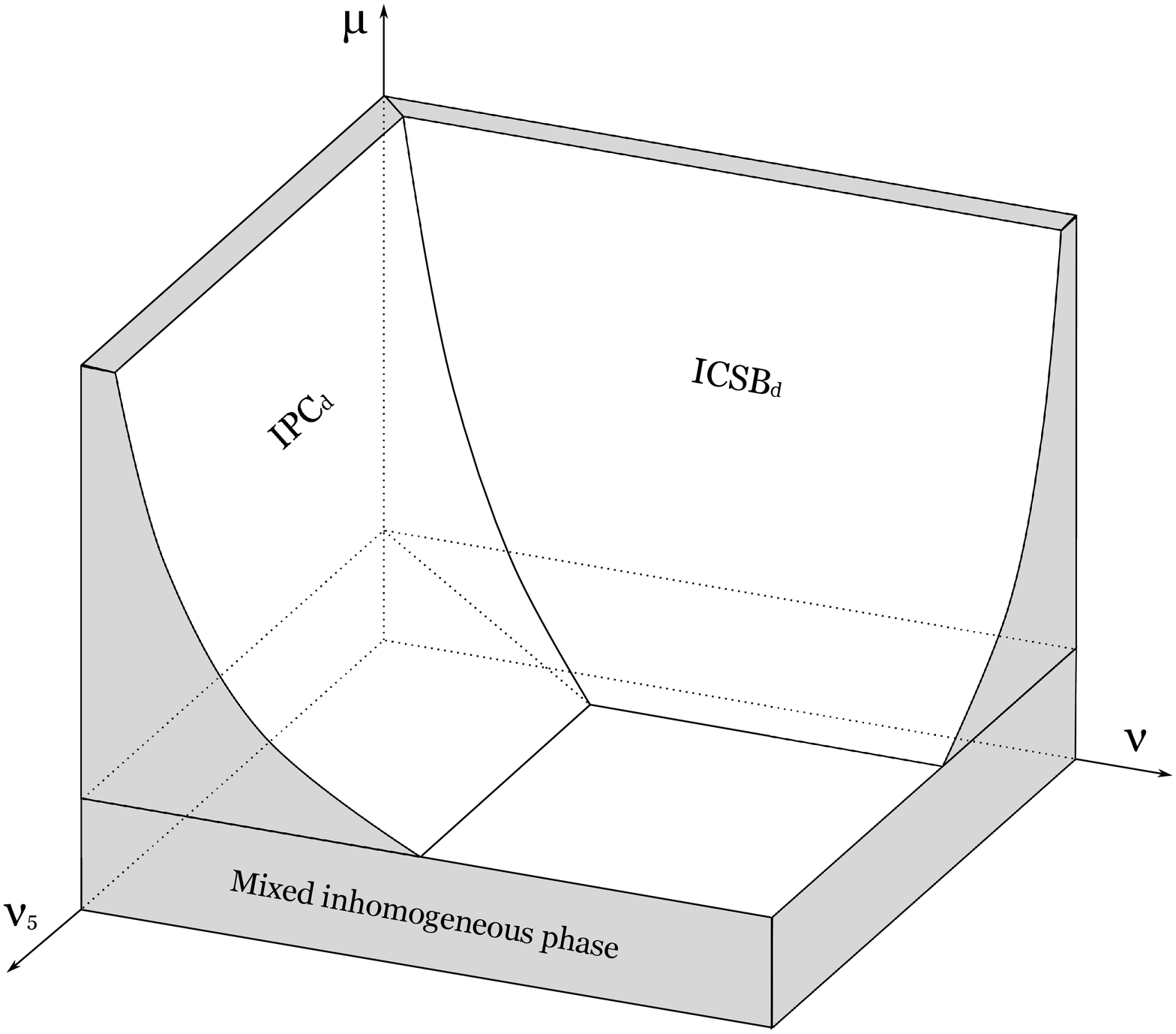}\\
\parbox[t]{0.45\textwidth}{\caption{{\bf The case of spatially inhomogeneous condensates}: The behavior of the coordinates $M_0,k_0,k'_0,\Delta_0$ of the GMP of the TDP (\ref{43}) as functions of $\nu_5$ for fixed $\mu=m$ and $\nu=0.1m$.}}\hfill
\parbox[t]{0.45\textwidth}{\caption{Schematic representation of the $(\nu_5,\nu,\mu)$-phase portrait of the model in the case of spatially inhomogeneous condensates. 
}}
\end{figure*}
Likewise at arbitrary fixed $\mu$ the $(\nu,\nu_5)$-phase diagram of the model is also self-dual. 
These conclusions are supported by Fig. 4. 
If $\mu< m/\sqrt{2}$, then the $(\nu,\nu_5)$-phase portrait is even simpler because at each point of it the ``Mixed inhomogeneous phase`` is realized. 

Now let us show how to construct the $(\nu_5,\mu)$-phase diagram of the model at arbitrary fixed value $\nu=A$ using duality transformation. 
 For example, to find the $(\nu_5,\mu)$-phase diagram at $\nu=0.1m$ we should start from the $(\nu,\mu)$-diagram at fixed $\nu_5=0.1m$ of Fig. 3 (left panel) and make 
 the replacement 
 : $\nu\leftrightarrow\nu_5$, IPCd$\leftrightarrow$ICSBd. 
 As a result 
 we obtain the phase diagram of Fig. 5. In a similar way one can dually transform Fig. 2 and Fig. 3 (right panel)  in order to find the $(\nu_5,\mu)$-phase diagrams at $\nu=0_+$ and $\nu\ge 0.2m$, respectively, etc. 

The behaviour of different order parameters 
are presented in Fig. 6. It is clear from the figure that in the critical point $\nu_5=0.1m$ there is a phase transition in the system from ICSBd phase 
to IPCd phase. 

The most general phase portrait ($\nu,\nu_5,\mu$ ) is presented schematically 
 at Fig. 7. As is easily seen from this figure 
 the phase diagram is self-dual. Moreover, it supports the above conclusion: the charged PC phenomenon can be realized in chirally asymmetric quark matter with nonzero baryon density.

\section{Summary and conclusions}

In this talk the phase structure of the NJL$_2$ model (1) with two
quark flavors is investigated in the large-$N_c$ limit in the presence
of baryon $\mu_B$, isospin $\mu_I$ and chiral isospin $\mu_{I5}$ chemical
potentials. 
Let us summarize some of the most interesting results obtained.

1) 
 The chemical potential $\mu_{I5}$ generates charged pion condensation in dense quark matter.

2) In this model inhomogeneous condensates are quite favoured compared to
homogeneous condensates. 

3) We demonstrated in the framework of the NJL$_2$ model (1), that 
in the leading order of the large-$N_c$ approximation there is duality
correspondence between CSB and charged PC phenomena. 

\bibliographystyle{ws-procs961x669}
\bibliography{ws-pro-sample}

\begin{thebibliography}{10}


\bibitem{njl}
Y. Nambu and G. Jona-Lasinio, Phys. Rev. D {\bf 112}, 345 (1961).








\bibitem{son}
D.T.~Son and M.A.~Stephanov, Phys.\ Atom.\ Nucl.\  {\bf 64}, 834 (2001);
M. Loewe and C. Villavicencio, Phys. Rev. D {\bf 67}, 074034
(2003); M. Loewe and C. Villavicencio, ``Pion stability in a hot dense media'', arXiv:1107.3859;
L. He, M. Jin, and P. Zhuang, Phys. Rev. D {\bf 71}, 116001 (2005);
D.C.~Duarte, R.L.S.~Farias and R.O.~Ramos,
  Phys.\ Rev.\  D {\bf 84}, 083525 (2011);
D.~Ebert, K.G.~Klimenko, A.V.~Tyukov and V.C.~.Zhukovsky,
  Eur.\ Phys.\ J.\ C {\bf 58}, 57 (2008).

\bibitem{eklim}
D. Ebert and K.G. Klimenko, J.\ Phys.\ G {\bf 32}, 599 (2006);
Eur.\ Phys.\ J.\  C {\bf 46}, 771 (2006).

\bibitem{ak}
J.O.~Andersen and T.~Brauner,
  Phys.\ Rev.\  D {\bf 78}, 014030 (2008);
J.O.~Andersen and L.~Kyllingstad,
 J.\ Phys.\ G {\bf 37}, 015003 (2009);
 Y.~Jiang, K.~Ren, T.~Xia and P.~Zhuang,
  ``Meson Screening Mass in a Strongly Coupled Pion Superfluid,''
  arXiv:1104.0094.

\bibitem{mu}
C.f.~Mu, L.y.~He and Y.x.~Liu,
  Phys.\ Rev.\  D {\bf 82}, 056006 (2010).






\bibitem{and}
J.O.~Andersen, W.R.~Naylor and A.~Tranberg,
  Rev.\ Mod.\ Phys.\  {\bf 88}, 025001 (2016).


\bibitem{gn}
D.J. Gross and A. Neveu, Phys. Rev. D {\bf 10}, 3235 (1974).

\bibitem{ft}
J.~Feinberg, Annals Phys.\  {\bf 309}, 166 (2004);
M.~Thies, J.\ Phys.\ A  {\bf 39}, 12707 (2006).



\bibitem{barducci}
A. Barducci, R. Casalbuoni, R. Gatto, M. Modugno, and G. Pettini, Phys. Rev. D {\bf 51}, 3042 (1995).

\bibitem{chodos}
 A.~Chodos, H.~Minakata, F.~Cooper, A.~Singh, and W.~Mao,
  Phys. Rev. D {\bf 61}, 045011 (2000);
 K.~Ohwa, Phys.\ Rev.\  D {\bf 65}, 085040 (2002).

\bibitem{thies}
V.~Schon and M.~Thies,
 Phys.\ Rev.\  D {\bf 62}, 096002 (2000);
A.~Brzoska and M.~Thies,
  Phys.\ Rev.\  D {\bf 65}, 125001 (2002).

\bibitem{coleman}
N.D. Mermin and H. Wagner, Phys.\ Rev.\ Lett. {\bf 17}, 1133
(1966); S. Coleman, Commun. Math. Phys. {\bf 31}, 259 (1973).

\bibitem{gubina}
  D.~Ebert, N.V.~Gubina, K.G.~Klimenko, S.G.~Kurbanov, V.C.~Zhukovsky,
  Phys.\ Rev.\ D {\bf 84}, 025004 (2011).


\bibitem{ek}
D.~Ebert and K.G.~Klimenko,
``Pion condensation in the Gross-Neveu model with nonzero baryon and isospin chemical potentials,''
  arXiv:0902.1861 [hep-ph].



  

\bibitem{ekk}
D. Ebert, T.G. Khunjua and K.G. Klimenko, Phys.\ Rev.\ D {\bf 94}, 116016 (2016)
[arXiv:1608.07688].

\bibitem{fukus}
K. Fukushima, D.E. Kharzeev and H.J. Warringa, Phys.Rev.D {\bf 78}, 074033 (2008).

\bibitem{andrianov}
 A.A.~Andrianov, D.~Espriu and X.~Planells,
  Eur.\ Phys.\ J.\ C {\bf 73}, 2294 (2013);
 Eur.\ Phys.\ J.\ C {\bf 74}, 2776 (2014);
R.~Gatto and M.~Ruggieri,
  Phys.\ Rev.\ D {\bf 85}, 054013 (2012);
 L.~Yu, H.~Liu and M.~Huang,
 Phys.\ Rev.\ D {\bf 90}, 074009 (2014);
L.~Yu, H.~Liu and M.~Huang,
 Phys.\ Rev.\ D {\bf 94}, 014026 (2016);
G.~Cao and P.~Zhuang,
  Phys.\ Rev.\ D {\bf 92}, 105030 (2015);
V.V.~Braguta and A.Y.~Kotov,
 Phys.\ Rev.\ D {\bf 93}, no. 10, 105025 (2016);
   M.~Ruggieri and G.~X.~Peng,
  arXiv:1602.05250 [hep-ph].

\bibitem{ms}
  V.A.~Miransky and I.A.~Shovkovy,
  Phys.\ Rept.\  {\bf 576}, 1 (2015).

\bibitem{grun}
G. Gruner, Rev. Mod. Phys. {\bf 66}, 1 (1994).

\bibitem{ferrel}
P. Fulde, R.A. Ferrell, Phys. Rev. A {\bf 135}, 550  (1964).

\bibitem{larkin}
A.I. Larkin, Y.N. Ovchinnikov, Zh. Eksp. Teor. Fiz. {\bf 47}, 1136  (1964) [Sov. Phys. JETP {\bf 20}, 762 (1965)].














\bibitem{weinberg}
S. Weinberg, ``The quantum Theory of Field II'', Cambridge Univ. Press, Cambridge, England, 1996.

\bibitem{fujikawa}
K. Fujikawa,   Phys.\ Rev.\  D {\bf 21}, 2848 (1980).

\bibitem{vafa}  
C.~Vafa and E.~Witten,
  Nucl.\ Phys.\ B {\bf 234}, 173 (1984).

\bibitem{miransky}
 E.V.~Gorbar, M.~Hashimoto and V.A.~Miransky,
  Phys.\ Rev.\ Lett.\  {\bf 96}, 022005 (2006);
J.O.~Andersen and T.~Brauner,
  Phys.\ Rev.\  D {\bf 81}, 096004 (2010);
C.f.~Mu, L.y.~He and Y.x.~Liu,
  Phys.\ Rev.\  D {\bf 82}, 056006 (2010).


\bibitem{Birkhoff}
G. Birkhoff and S. Mac Lane, ``A Survey of Modern Algebra``, New York:
Macmillan, 1977.



\end{thebibliography}



\end{document}